\documentclass{llncs}
\usepackage{amssymb}
\usepackage{graphicx}
\usepackage{color}
\usepackage{pstricks,pst-node,pst-tree}
\title{Towards a better insight of RDF triples Ontology-guided Storage system abilities}%\texttt{roStore}: an RDF triples Ontology-guided Storage system}
\author{Olivier Cur\'e\inst{1}, David Faye\inst{1,2}, Guillaume Blin\inst{1}}
\institute{Universit\'{e} Paris-Est,  LIGM - UMR CNRS 8049, France\\
\email{\{ocure, gblin\}@univ-mlv.fr}
\and 
Universit\'e Gaston Berger de Saint-Louis, LANI, S\'en\'egal\\
\email{dfaye@igm.univ-mlv.fr}
}

\begin{document}

\maketitle
\begin{abstract}
The vision of the Semantic Web is becoming a reality with billions of \texttt{RDF} triples being distributed over multiple 
queryable endpoints (e.g. Linked Data). 
Although there has been a body of work on \texttt{RDF} triples persistent storage, 
it seems that, considering reasoning dependent queries, the problem of providing an efficient, in terms of performance, scalability and 
data redundancy, partitioning of the data is still open. 
In regards to recent data partitioning studies, it seems reasonable to think that
data partitioning should be guided considering several directions (e.g. ontology, data, application queries).
This paper proposes several contributions: describe an overview
of what a roadmap for data partitioning for \texttt{RDF} data efficient and persistent storage should contain, present some preliminary
results and analysis on the particular case of ontology-guided (property hierarchy) partitioning and finally introduce a set of semantic query rewriting rules
 to support querying \texttt{RDF} data needing \texttt{OWL} inferences.
% In this work, we take benefit of recent papers proposing a vertically-partition approach implemented on a column-oriented relational database management system 
% and extend it with a novel approach enabling to represent less relations, thus requiring less joins in practical queries.
% This extension uses the fact that properties are first-class citizens in the RDF model to propose a particularly efficient solution when inferences are performed on property hierarchies.
% Another contribution of this paper is to propose a set of semantic query rewriting rules to improve query performance by reasoning over the ontology schema of
% the RDF triples.
% We provide an experimental evaluation on synthetic databases which emphasizes the relevance of our two contributions in efficient RDF triples storage.

\end{abstract}

\section{Introduction}

%%% Lots of RDf data available
% The vision of the Semantic Web, as proposed by
% \cite{Berners-lee2001Semanticweb}, is becoming a reality with billions of
% \texttt{RDF} triples and hundreds of \texttt{RDFS/OWL} ontologies being
% distributed over multiple queryable endpoints (e.g. Linked data).
% These endpoints can generally be queried using the \texttt{SPARQL} query
% language \cite{sparql08} or in a programmatic manner via one of the many
% available \texttt{RDF APIs} \cite{jena}. Practically, these queries usually 
% require some form of reasoning, a feature not natively supported by the 
% current \texttt{SPARQL} W3C's recommendation. 

The generally encountered use of ontologies consists in performing data inferences 
and validation using a Semantic Web compliant reasoner. The corresponding reasoning
mechanism can be used to generate a set of queries executed over the appropriate 
data sets. For example, this approach was designed in a medical application 
\cite{Cure:2005:SDM:2101422.2101472} where inferences on chemical molecules were needed to 
highlight contra indications, side effects of pharmaceutical products. 
As mentioned in \cite{Cure:2005:SDM:2101422.2101472},
results of queries with both inference on property (\textit{i.e.} \texttt{rdf:property}) and concept 
(\textit{i.e.} \texttt{rdf:class}) hierarchies are required by the 
application as well as by data quality or data exchange external tools.

In regards to large ontologies (e.g. \texttt{OpenGalen} or \texttt{SNOMED} in the medical domain) and data sets (e.g. Linked Data),
 providing efficient performances to reasoning dependent queries is an important
issue. We believe that to enable efficient response time to such queries, one has to
give a special attention to the storage system associated to the triples.
In fact, \texttt{RDF} is basically a data model and its recommendation does not guide
to a preferred storage solution. The related work about \texttt{RDF} data management systems 
can be subdivided into two categories: the ones involving a mapping to a Relational DataBase 
Management System (RDBMS) and the ones that do not. In this paper, we do not focus on the latter one.

\begin{figure}
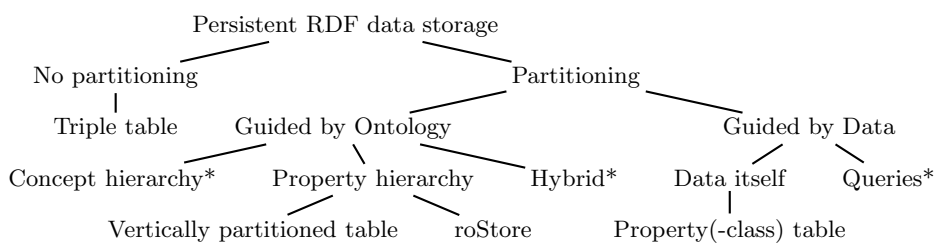

\pstree[nodesep=2pt,levelsep=5ex]{\TR{Persistent RDF data storage} }{
    \pstree{ \TR{No partitioning} }{
      \TR{Triple table}
     }
    \pstree{ \TR{Partitioning} }{
      \pstree{ \TR{Guided by Ontology} }{
	\TR{Concept hierarchy*}
	\pstree{ \TR{Property hierarchy} }{
	    \TR{Vertically partitioned table}
	    \TR{roStore}
	}        
        \TR{Hybrid*}
      }
      \pstree{ \TR{Guided by Data} }{
	  \pstree{ \TR{Data itself} }{
	      \TR{Property(-class) table}
	   }
           \TR{Queries*}
      }
    }
}\caption{Physical organization of \texttt{RDF} data based on partitioning. (*) no known study yet.}\label{phys_org}
\end{figure}

A set of techniques have been proposed for storing \texttt{RDF} data in
relational databases. Several research groups think that this is likely the
best performing approach for their persistent data store, since a great amount
of work has been done on making relational systems efficient, extremely scalable
and robust. Efficient storage of \texttt{RDF} data has already been discussed in
the literature with different physical organization techniques based on partitioning (cf. Figure \ref{phys_org}).

On one hand, there exists tools such as Sesame \cite{Broekstra:2002:SGA:646996.711426}, Jena \cite{Wilkinson03efficientrdf}, 
Oracle \cite{Chong:2005:ESR:1083592.1083734} and 3store \cite{3store2003} which rely on a straightforward 
mapping of \texttt{RDF} into an RDBMS -- called \textit{triple table} approach. Each 
\texttt{RDF} statement of the form $(subject,$ $predicate,$ $object)$ is stored as an entry of
 one large table  with a three-columns schema (\textit{i.e.} a column for each the \textit{subject},
\textit{predicate} and \textit{object}). Indexes are then added for each of the columns in
order to make joins less expensive. However, since the collection of triples are stored in one 
single table, the queries may be very slow to execute. Indeed when the number of
triples scales, the table may exceed memory size (inducing costly disk-RAM transfers). Nevertheless, simple \emph{statement-based} 
queries can be satisfactorily processed by such systems, although they do not represent the 
most important way of querying \texttt{RDF} data. Still, this storage system scales poorly since 
complex queries with multiple triple patterns require many self-joins
over this single large table as pointed out in \cite{Wilkinson03efficientrdf,Weiss:2008:HSI:1453856.1453965,Kolas_efficientlinked-list}.

%\item Efficienciy = Partioning in order to get better response time by parsing less data
%\item Inherent paradox : The more partioning = The less efficient for updates
Whereas this specific approach does not use partitioning at all, on the other hand, some recent research  highlighted two efficient main trends 
depending on the information one uses to guide the partitioning: guided by (1) the underlying 
ontology or (2) the data itself. Intuitively, one would expect that a well suited data partitioning will induce 
a better response time to queries (at least \textsc{select} ones). Indeed, data partitioning will allow 
queries to be made on smaller sets of entries which, given an adapted \texttt{RDF} data clustering, should 
be faster. The counterpart of this storage system will be some possible worst performance for
data updates. 
% On the following, we will mention both known results and new tracks for future work for those two approaches.

\section{The \texttt{roStore} approach}
\label{ROStore}

As a first step to an efficient \texttt{RDF} storage road map, we propose an intermediate ontology-guided approach -- namely 
\texttt{roStore} -- which lies between the two extremes: triple and vertically partitioned tables. The aim of this approach is to 
try to analyse the efficiency of a compromise approach where less partitions are used. Intuitively, 
such physical organization will take benefits of requiring less joins in practical queries and with less risk of unmappable table in memory.

As already mentioned, we believe that there should not be a unique generic solution to \texttt{RDF} storage and that depending
on the data itself, the underlying ontology, application queries, better performance may be obtained by considering alternative and several
dedicated approaches. The major aim of \texttt{roStore} is to provide some clue of this belief. We will demonstrate that \texttt{roStore} is one of them and may, in specific cases, induce better efficiency.
In this context, we consider \texttt{roStore} as \textbf{one among other} interesting physical organizations based on property hierarchy that should be
present in the \texttt{RDF} storage road map. 

Our storage approach derives from the vertically partitioned one and extends this last by putting back together into a single table 
data related to a \textit{top-property} of a property hierarchy. Given a hierarchy, we say that a \textit{predicate} is a top-property if it
is only an \texttt{rdf:subPropertyOf} of itself. For each such top-property $P^{T}$, a three-columns table is created by 
(1) merging all the two-columns tables corresponding to \textit{predicates} being \texttt{rdf:subPropertyOf} $P^T$ and (2) adding a third 
column indicating from which \textit{predicate} the entry \textit{(subject, object)} was retrieved (cf. Figure \ref{storageComp}). 

Let us first notice that, providing with this definition, any \textit{predicate} that is not an \texttt{rdf:subPropertyOf} of a top-property
will still be stored in a two-columns table. This induces an insignificant expense of the space complexity of this novel approach.
Moreover, in case of a cyclic \textit{property} hierarchy, all \textit{predicates} are necessarily all semantically equivalent.
Hence selecting a single canonical \textit{predicate} and rewriting triples accordingly is sufficient.
Despite the fact that considering top-property seems to be the most natural, one may, depending on the topology of the hierarchy, define other physical organizations inducing better 
performance too for specific cases. Our preliminiary results demonstrate that an evolutive physical organizations guided by the queries may be efficient. The main 
impact of merging some tables is obtaining better performance of queries requiring joins over \textit{predicates} belonging to the
same ``sub-hierarchy`` of the property hierarchy. This is typically the case when one wants to retrieve all the information concerning 
a family of \textit{predicates} of the property hierarchy; since they will be quite related. In the following, we will denote by \texttt{vpStore} 
(resp. \texttt{roStore}) the vertically partitioned (resp. our) approach.

% In \texttt{swStore} and \texttt{RIBStore}, a two columns relation is created for properties of the \texttt{RDF} and \texttt{RDFS} vocabularies, 
% \textit{e.g.} rdf:type, rdfs:subClassOf.
% And for all other properties of the ontology, we apply the following approach: (1) if a property is not part of a property hierarchy, 
% generate a relation as in \texttt{swStore}. That is a two columns relation with subject and object attributes and the property name as relation 
% name; (2) otherwise, for all properties in a property hierarchy, generate a single relation with the name of the top property of the hierarchy 
% as relation name and a three columns pattern (subject, object and property). Each tuple of this relation contains the name of the property of 
% the corresponding triple in the property column.

\textbf{Example 1:}
Let us consider a small data set (Figure \ref{storageComp}b) defined over a given property hierarchy (Figure \ref{storageComp}a). 
With \texttt{vpStore}, the triples would be distributed over six different tables as displayed in Figure \ref{storageComp}c.
Comparatively, in \texttt{roStore}, one obtains only two different tables (Figure \ref{storageComp}d): a single relation named after the 
top-property \texttt{pa} and a relation named after the property \texttt{pf}.

\begin{figure}[ht!]
\centering

%\begin{flushleft}
\begin{minipage}[t]{4cm}
(a) property hierarchy: %\\
%\texttt{pb} $\sqsubseteq$
%\texttt{pa}, \\\texttt{pc} $\sqsubseteq$ \texttt{pa}, \\\texttt{pd}
%$\sqsubseteq$ \texttt{pc}, \\\texttt{pe} $\sqsubseteq$ \texttt{pc}

pf\hspace*{0.2cm}\pstree[nodesep=2pt,levelsep=5ex]{\TR{pa} }{
      \TR{pb}
      \pstree{ \TR{pc} }{
	  \TR{pd}
           \TR{pe}
      }    
}
\end{minipage}
%\hfill
\begin{minipage}[t]{3cm}
(b) RDF triples \\
\newline
\begin{tabular}{|c|c|c|}
 \hline
sub.&prop.&obj.\\
\hline
a&pa&b\\
\hline
c&pc&d\\
\hline
e&pb&f\\
\hline
a&pf&d\\
\hline
g&pe&h\\
\hline
\end{tabular}
\end{minipage}
%\hfill
\newline
\newline
\begin{minipage}[t]{5cm}
(c) \texttt{vpStore}\\
\newline
\begin{tabular}{|c|c|}
 \hline
sub.&obj.\\
\hline
a&b\\
\hline
\end{tabular}
\hfill
\begin{tabular}{|c|c|}
 \hline
sub.&obj.\\
\hline
e&f\\
\hline
\end{tabular}
\hfill
\begin{tabular}{|c|c|}
 \hline
sub.&obj.\\
\hline
c&d\\
\hline
\end{tabular}
\\\texttt{pa} relation \hfill\texttt{pb} relation \hfill\texttt{pc} relation\\
\begin{tabular}{|c|c|}
 \hline
sub.&obj.\\
\hline
&\\
\hline
\end{tabular}
\hfill\begin{tabular}{|c|c|}
 \hline
sub.&obj.\\
\hline
g&h\\
\hline
\end{tabular}\hfill\begin{tabular}{|c|c|}
 \hline
sub.&obj.\\
\hline
a&d\\
\hline
\end{tabular}
\\
\texttt{pd} relation \hfill\texttt{pe} relation\hfill\texttt{pf} relation\\
\end{minipage}
\hfill
\begin{minipage}[t]{5cm}
(d) \texttt{roStore}\\
\newline
\begin{tabular}{|c|c|c|}
 \hline
sub.&obj.&prop.\\
\hline
a&b&pa\\
\hline
c&d&pc\\
\hline
e&f&pb\\
\hline
g&h&pe\\
\hline
\end{tabular}
\hfill
\begin{tabular}{|c|c|}
 \hline
sub.&obj.\\
\hline
a&d\\
\hline
\end{tabular}
\\
\texttt{pa} relation \hfill\texttt{pf} relation\\
\end{minipage}
%\end{flushleft}
\caption{Storage comparison of \texttt{vpStore} and \texttt{roStore}}\label{storageComp}
\end{figure}
Thus, if we consider an ontology consisting of \texttt{n} (e.g. 2 in our example) property hierarchies with an average of 
\texttt{k} (e.g. 3 in our example) properties in each hierarchy, the \texttt{roStore} approach will store 
\texttt{k} times less tables than a \texttt{vpStore} approach. Moreover, with this approach it is very unlikely 
to generate tables with no tuples (e.g. \texttt{pd} with \texttt{vpStore} in Example 1).
Moreover, the set of tuples stored is the same as in \texttt{vpStore} and only their distribution over database tables is modified (\textit{i.e.} physical organization).

We now consider the following query: one wants to retrieve all \textit{objects} involved in a triple with a \textit{predicate} of 
the \texttt{pa} hierarchy. Considering \texttt{vpStore}'s physical design, the following \texttt{SQL} query is needed:

\medskip

\texttt{SELECT object FROM pa UNION (SELECT object FROM pb UNION (SELECT object FROM pc UNION (SELECT object FROM pd UNION (SELECT object FROM pe))));}

\medskip

while the same query is answered far more efficiently considering \texttt{roStore}'s physical design with:

\medskip

\texttt{SELECT object FROM pa;}

\medskip

Such example highlights the kind of (1) reasoning dependent queries and (2) corresponding improvement one can obtain by using 
an intermediate physical organization such as \texttt{roStore} over \texttt{vpStore} when property hierarchies are present in the ontology. 

In order to analyse more deeply the corresponding efficiency of \texttt{roStore} approach, we will first compare it to the \texttt{vpStore} approach 
on the LUBM benchmark and on some specific queries that highlight limits of vertically partitioning. In this work, as a first contribution, 
we focus only on \textsc{select} queries. We are currently investigating the possibly negative impact of partitioning on \textsc{update}
queries. As far as we went on this track, it seems that this impact is reasonable.
%  and should not be a priceless cost to pay to get efficiency. 
First, we will discuss how to take benefits of the ontology based structure of the data without the needs of heavy inference mechanisms.
 
Indeed, compared to classical table schema, the ontology is far more meaningfull and can thus be used to enhance the performance even 
without needing knowledge inference. We propose an efficient use of Semantic Query Rewriting (SQR) adaptable to and usable by most of the data 
storage approaches. Our semantic query rewriting aims are, first, to guarantee the exhaustiveness of results returned when requiring data 
that should include \texttt{rdf:subClassOf} and \texttt{rdf:subPropertyOf}; and, a query validation mechanism simply based on the domain and range 
information related to \textit{predicates}, \textit{i.e.} resp. \texttt{rdfs:domain} and \texttt{rdfs:range}. 
One has to note that the mechanisms we propose are not really needing heavy reasoning nor data inference mechanisms. Indeed, they can be considered as 
an efficient use of the right-away available information of the ontology.

The semantic aspect of this rewriting is provided by a thorough usage of the \texttt{OWL} entailment mechanism, on one hand, to detect if the answer set of a query will be empty or not, on the other hand,
to optimize query in order to guarantee exhaustiveness of the solution returned.
The rules can be decomposed into two sets: (i) a set of rules, denoted \texttt{subsume}, dealing with concept and property subsumptions; (ii) a set of rules, denoted \texttt{propertyCheck}, dealing with the 
\texttt{rdfs:range} and \texttt{rdfs:domain} of a given \textit{predicate}.
The rules processed by the \texttt{subsume} procedure are using the \texttt{OWL} inferences to compute all the sub-concepts 
(resp. sub-properties) of a given concept (resp. property). In fact, the query studied in Example 1 was already using the \texttt{subsume} procedure. 

\textbf{Example 2:}
Consider that the \texttt{rdfs:range} of the \textit{predicate} \texttt{pb} of Example 1 is of \texttt{rdf:type} \texttt{ClassA} which is the 
top-concept in the following concept hierarchy:

$$
\texttt{ClassC} \sqsubseteq \texttt{ClassA},\; \texttt{ClassB} \sqsubseteq \texttt{ClassA}\; and\; \texttt{ClassC} \sqsubseteq \neg \texttt{ClassB}
$$ 

That is \texttt{ClassA} has two sub-concepts which are disjoint. Consider a query asking for all \textit{subjects} and \textit{objects} of triples 
where \texttt{pb} is the \textit{predicate} and all \textit{subjects} belong to the \texttt{ClassA} hierarchy. Using \texttt{subsume}, the query 
can be translated in the following \texttt{SQL} query:

\medskip

\texttt{SELECT subject, object FROM pa, type WHERE type.subject =\\ 
pa.subject AND pa.property = 'pb' AND type.object IN \\('ClassA','ClassB', 'ClassC');}
\medskip

Thus this approach enables to generate a singe \texttt{SQL} query whatever the size of the concept hierarchy is.
Note that it also applies to the property hierarchy.

The rules of \texttt{propertyCheck} are being processed as follows: first the \texttt{SPARQL} query is parsed and for each \textit{predicate} 
explicitly mentioned in the query with a typed (\texttt{rdf:type}) \textit{subject} or \textit{object}, we store a structure containing 
the \textit{predicate} name and the 
\texttt{rdf:type} of the \textit{subject} and/or \textit{object}. Then for each \textit{subject} (resp. \textit{object}) in the structure, we 
search if there is a direct or indirect (via subsumptions) correspondence with the type of the \texttt{rdfs:domain} (resp. \texttt{rdfs:range}) 
defined in the ontology for this property.

\textbf{Example 3:}
Let us consider the property hierarchy of Figure \ref{ibuprofen}, dealing with contra indications and the 
corresponding \texttt{roStore} organization.

\begin{figure}
\centering

\pstree[nodesep=2pt,levelsep=5ex]{\TR{contraIndication} }{
      \TR{diseaseContraIndication}
    \TR{moleculeContraIndication}
	\TR{stateContraIndication}
      
}\\
\vspace*{0,5cm}\begin{tabular}{|l|l|l|}
\hline
\textbf{subject} & \textbf{object} & \textbf{property}\\
\hline
Ibuprofen & Ticlopidin & moleculeContraIndication\\
\hline
Ibuprofen & Clopidrogel & moleculeContraIndication\\
\hline
Ibuprofen & Breast feeding & stateContraIndication\\
\hline
Ibuprofen & Pregnant & stateContraIndication\\
\hline
Ibuprofen & Hypertensive heart & diseaseContraIndication\\
\hline
\end{tabular}
\caption{Sample of the contraIndication relation}\label{ibuprofen}
\end{figure}

Moreover, consider the following ontology axioms: (1) \texttt{rdf:range} of \texttt{disease} \texttt{ContraIndication} is an instance of 
the \texttt{Disease} concept, (2) \texttt{Disease} $\sqsubseteq$ \texttt{Top},
(3) \texttt{Molecule} $\sqsubseteq$ \texttt{Top} and (4) \texttt{Disease} $\sqsubseteq \neg$ \texttt{Molecule}. 
Intuitively, axioms (2) to (4) state that the \texttt{Disease} and \texttt{Molecule} concepts are a sub-concept ot the 
\texttt{Top} concept and are disjoint. Consider the following \texttt{SPARQL} query:

\medskip

\texttt{SELECT ?s ?o WHERE \{?s :diseaseContraIndication ?o.\\ ?o rdf:type :Molecule.\}}

\medskip
 
which asks for \textit{subjects} and \textit{objects} involved in triples where the \textit{predicate} is \texttt{diseaseContraIndication} and 
the \textit{object} has a \texttt{rdf:type} \texttt{Molecule}. Clearly the answer set to this query is empty since the \texttt{rdf:domain} of 
the \textit{predicate} can not be a \texttt{Molecule} in this ontology.

\textbf{Example 4:}
Consider the following query in the context of Example 3:

\medskip

\texttt{SELECT ?s ?o WHERE \{?s :diseaseContraIndication ?o.\\ ?o rdf:type :Disease.\}}

\medskip

The query is satisfiable since there is a model where its answer set is not empty. Anyhow, the query can be optimized.
In fact, it is not necessary to check the \texttt{rdf:type} of the \textit{object} because it corresponds exactly to the one defined as 
\texttt{rdf:range} in the ontology. Thus this query is rewritten in:

\medskip

 \texttt{SELECT ?s ?o WHERE \{?s :diseaseContraIndication ?o.\}}

\medskip

which once translated into \texttt{SQL} does not require any join and will thus perform far more efficiently than the orginal query.
Note that this simplification does not work for property with multiple-range/domain.
Those examples demonstrate that it is worth to efficiently use the basic knowledge available directly in the concept and property hierarchies.

\section{Evaluation}\label{sec_eval}

\subsection{Experimental settings}
All our experiments have been conducted on four synthetic databases.
They all have been generated from the Lehigh University Benchmark (LUBM) \cite{DBLP:journals/ws/GuoPH05} which has been developed to facilitate
the evaluation of Semantic Web repositories in a standard and systematic way.
The \texttt{RDF} data sets generated with LUBM all commit to a single realistic ontology dealing with the university domain.
This ontology is composed of 43 concepts, 25 object properties (\textit{i.e.} relating objects to objects) and 7 data type properties (relating objects to literals).

% Since we are interested in inferences using concept and property hierarchies, our queries will take advantage of the 36 sub class and 5 sub property axioms. 
% Some of our queries imply to infer on subclass axioms. For this purpose, we have selected the hierarchy of the $Person$ class from which we present and extract:
% $AssociateProfessor \sqsubseteq Professor \sqsubseteq Faculty \sqsubseteq Employee \sqsubseteq Person$, $AdministrativeStaff \sqsubseteq Employee$ and $
% Faculty \sqsubseteq \neg AdministrativeStaff$.
% The sub property axioms we are using the most in our evaluation (queries 1 to 6) concern the membership hierarchy which is classified as follows : $headOf \sqsubseteq worksFor \sqsubseteq memberOf$.
% Surprisingly, these 3 object properties are underspecified since no domain and range are provided.
% Since we also want to evaluate our query rewriting technique, queries 7 to 9 are using the $teacherOf$ property whose domain and range are clearly defined.

This ontology serves as the schema underlying the four data sets we have created. This is an important requisite for our evaluation since our set of queries will 
be executed on all data sets in order to provide information on scalability issues. Table \ref{syntheticDB} summarizes the main characteristics of these data sets 
in terms of overall number of triples, number of concept and property instances.

\begin{table}
\centering
\caption{Synthetic data sets}
\label{syntheticDB}
\begin{tabular}{|c|c|c|c|c|} 
\hline
DB name & \# Universities & \# Concept instances & \# Property instances & \# Triples\\ 
\hline
lubm1 & 1 & 15195 & 60859 & 100868\\ 
\hline
lubm2 & 2 & 62848 & 189553 & 236336\\
\hline
lubm5 & 5 & 114535 & 456137 & 643435\\
\hline
lubm10 & 10 & 263427 & 1052895 & 1296940\\
\hline
\end{tabular}
\end{table}

The \texttt{RDF} data sets are later translated into the different physical organization models we would like to evaluate.
They are decomposed into the two main approaches \texttt{vpStore} and \texttt{roStore}. 
In order to emphasize the efficiency of our solution on queries needing reasoning, we had to test these settings in a context similar to \cite{Abadi}.
More precisely, we evaluated each approach on a row store and a column store RDBMS. This yields the four following approaches: 
\texttt{vpStore} resp. on a row (\textbf{vpRStore}) and column (\textbf{vpCStore}) store and \texttt{roStore} resp. on a row 
(\textbf{roRStore}) and column (\textbf{roCStore}) store. Hence a total of sixteen databases are generated (each data set is 
implemented on each physical approach).

We have selected postgreSQL and MonetDB as the RDBMS resp. for the row-oriented and the column-oriented databases.
We retained MonetDB instead of C-store (the column store used for evaluation in \cite{Abadi}) essentially due to (1) the lack of maintenance 
of the latter one, (2) the open-source licence of MonetDB and (3) the fact that MonetDB is considered state of the art in column-oriented 
databases. The tests were run on MonetDB server version 5 and postgreSQL version 8.3.1. 
The benchmarking system is an Intel Pentium 4 (2.8 GHz) operated by a Linux Ubuntu 9.10, with 512 Mbytes of memory, 1MB L2 cache
and one disk of 60 Gbyte spinning at 7200rpm. The disk can read cold data at a rate of approximatively 55MB/sec.

For the \texttt{vpRStore}, there is a clustered B+ tree index on the \textit{subject} and an unclustered B+ tree on the \textit{object}.
Similarly, for the \texttt{roRStore}, a clustered B+ tree index is created on the \textit{property} column and unclustered B+ trees on the 
\textit{subject} and \textit{object}. As noted in \cite{Sidirourgos08}, MonetDB does not include user defined indices. Hence, we relied on 
the ordering of the data on \textit{property}, \textit{subject} and \textit{object} values. More precisely, any two columns table of 
\texttt{roCStore} and \texttt{vpCStore} is ordered on \textit{subject} and \textit{object}; while any three columns table (of \texttt{roCStore})
is ordered on \textit{property}, \textit{subject} and \textit{object}.

Our evaluation contains fifteen queries out of which eleven are coming from the LUBM benchmark and four tackling the LUBM ontology to evaluate 
some particular aspects of \texttt{roStore}.
% Indeed, out of the fourteen LUBM queries, we have selected only the queries that rely on \texttt{RDFS} reasoning.
% That is queries involving \texttt{OWL}, namely, queries \#11 (transitive aspect of the \texttt{subOrganizationOf} \textit{predicate}), \#12 (requiring \texttt{DL} realization inference 
% procedure) and \#13 (involving the inverse of a \textit{predicate}) were removed from the evaluation query set.
An interesting aspect using LUBM Benchmark queries is that do not aim to emphasize on the performances of a given storage model. 
Moreover, these queries tackle a wide range of possibilities 
on volume of input (number of tuples retrieved) and selectivity rate (\textit{i.e.} number of conditions in the \texttt{WHERE} clause of a query).  
Among the eleven evaluated queries, three do not require any form of reasoning (\#1, \#2 and \#14) and the eight remaining queries can be divided in two groups whether 
they are involving reasoning on the concept hierarchy (\#3,\#4,\#6,\#7,\#9,\#10) or both concept/property hierarchies (\#5,\#8). 
We now present the purpose of each of these queries:

\textbf{Q1:} retrieves instances of the \texttt{GraduateStudent} class who have taken the course \texttt{http://www.Department0.University0.edu/GraduateCourse0}.

\textbf{Q2:} retrieves three instances of respectively the \texttt{GraduateStudent}, \texttt{University} and \texttt{Department} concepts for 
those students that are member of a department, this department is a sub-organization of a University and this student has an undergraduate degree 
from this university.

\textbf{Q3:} selects all kinds of publications which have been authored by a given assistant professor.

\textbf{Q4:} retrieves all kinds of professors, their name, email address and telephone number for those professors working for a given 
department. 

\textbf{Q5:} the result contains instances of the \texttt{Person} concept hierarchy for those persons that are related to a given department by
either the \texttt{memberOf}, \texttt{workingFor} or \texttt{headOf} properties. 

\textbf{Q6:} displays URIs of instances of the \texttt{Student} concept hierarchy.

\textbf{Q7:} retrieves instances of all kind of students and all kinds of courses for courses that are related by the \texttt{takesCourse} property 
for those courses that are taught by a given professor.

\textbf{Q8:} displays instances of all kinds of students with their email addresses and department instances of a given university
 these students are member of.

\textbf{Q9:} the retrieved dataset contains instances of the \texttt{Student}, \texttt{Faculty} and \texttt{Course} concept hierarchies for 
those students that are advised by faculties, have taken some courses taught by those faculties.

\textbf{Q10:} selects instances of all the \texttt{Student} class hierarchy who have taken a given course.

\textbf{Q14:} selects all undergraduate students.

We have introduced \textbf{Q15} to emphasize \texttt{roStore} performances when values are needed for a property hierarchy. In fact, it 
 retrieves all \textit{subjects} involved in triples where the \textit{predicate} is one of the properties of the \texttt{memberOf} property hierarchy,
 i.e. \texttt{memberOf}, \texttt{headOf} and \texttt{worksFor}. This query is similar to Q5 but does not refer to any concepts.

Finally, the following three queries aim to highlight the efficiency of our SQR approach.
\textbf{Q16:} selects the \textit{subject} and \textit{object} in triples where the \textit{predicate} is \texttt{teacherOf} and \textit{subject} is 
of \texttt{rdf:type} AdministrativeStaff. This query returns an empty answer set since the \texttt{rdfs:domain} of \texttt{teacherOf} is the \texttt{Faculty}
 concept which is disjoint with \texttt{AdministrativeStaff}. In the next section, we confront the performances of this query to the simple 
detection of unsatisfiability of our SQR solution.

\textbf{Q17:}  selects the \textit{subject} and the \textit{object} in triples where the \textit{predicate} is \texttt{teacherOf} and \textit{subject} 
is of \texttt{rdf:type} \texttt{Faculty}. This query requires a join.

\textbf{Q18:} has the same purpose as Q17 but exploits one of our rewriting rules to improve its performances. In fact, the join in Q17 is not 
necessary if one knows that the \texttt{rdfs:domain} of \texttt{teacherOf} is the concept \texttt{Faculty}.

In the experiments, we will store the LUBM ontology in main-memory and perform reasoning using the Jena framework. We provide more details concerning the experimental 
settings and results on the following web site: \\ \texttt{http://sites.google.com/site/wwwrostore}.

\subsection{Experimental results}
The results presented in this section correspond to the average of 5 hot runs (i.e. repeated runs of the same query without stopping the DBMS) of real time
 (i.e. execution time of a query defined as the wall clock passed between the server receiving the query and before returning the results to the client) executions. 
All performance times, except for query Q16 and Q17, include the time needed to perform the inferences.
Finally, in order to highlight the differences in terms of performances between the various approaches, we either present the results in bar or line diagrams.

\noindent\textbf{Analysis of Q1.} Not surprisingly, column stores outperform row stores. Indeed, the results will only contain a unique column which 
will clearly benefit column store advantage. Moreover, since the query does not involve sub-properties, 
the performances of \texttt{vpStore} and \texttt{roStore} are quite similar.

\noindent\textbf{Analysis of Q2.} This time, the row stores are more efficient than the column ones. The results require, in this case, to retrieve two columns of three tables, hence 
in a row store both columns will be transfered from the hard drive to main memory in a single step while two transfers will be needed for column stores.
Moreover, two out of these three tables corresponds to \textit{predicates} being part of a group of related predicates in the property hierarchies, namely 
\texttt{memberOf} and \texttt{undergraduateDegreeFrom}. Since we voluntarily decided to perform no inferences on these two \textit{predicates}\footnote{since it will be 
specifically considered in query Q15.} (\textit{i.e.} not including sub-properties in the query), it is not surprising that \texttt{vpStore} 
outperforms \texttt{roStore} since each \textit{predicate} corresponds in the \texttt{vpStore} to a table.

\noindent\textbf{Analysis of Q3.} Despite the fact that this query has a similar structure as Q1 (\textit{i.e.} only two triples are present in the \texttt{WHERE} clause), it requires to retrieve 
all concepts of a wide hierarchy. Due to the ordered organization of tuples, column stores outperform row ones which rely on indices and on a less effective I/O transfers. 
In a similar manner to Q1, the difference between \texttt{vpStore} and \texttt{roStore} is not significant.

\begin{figure}[ht]
\begin{minipage}[b]{0.5\linewidth}
\centering
\includegraphics[scale=.6]{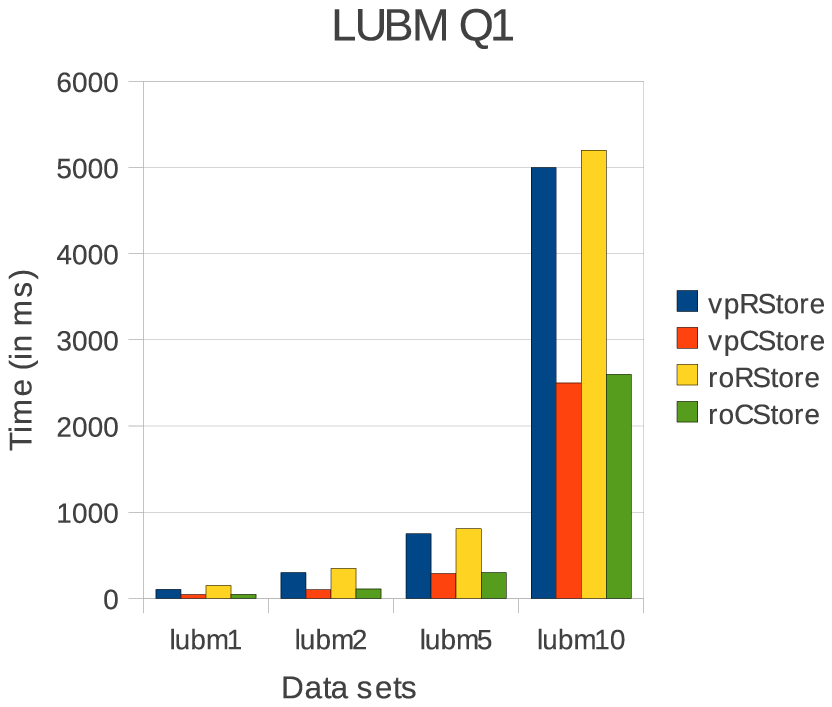}
\caption{Performance results for Q1}
\label{q1}
\end{minipage}
\hspace{0.1cm}
\begin{minipage}[b]{0.5\linewidth}
\centering
\includegraphics[scale=.6]{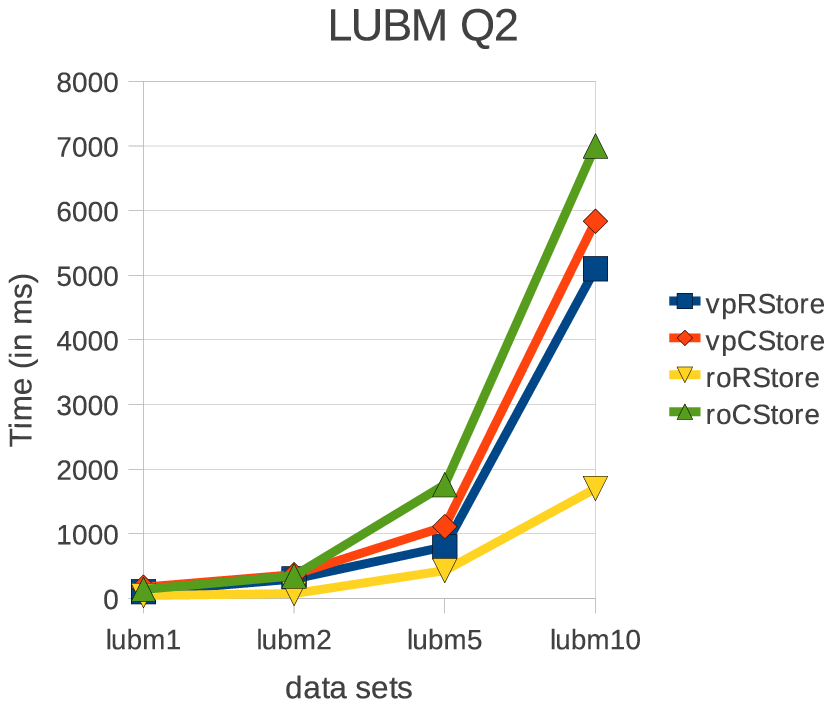}
\caption{Performance results for Q2}
\label{q2}
\end{minipage}
\end{figure}

\begin{figure}[ht]
\begin{minipage}[b]{0.5\linewidth}
\centering
\includegraphics[scale=.6]{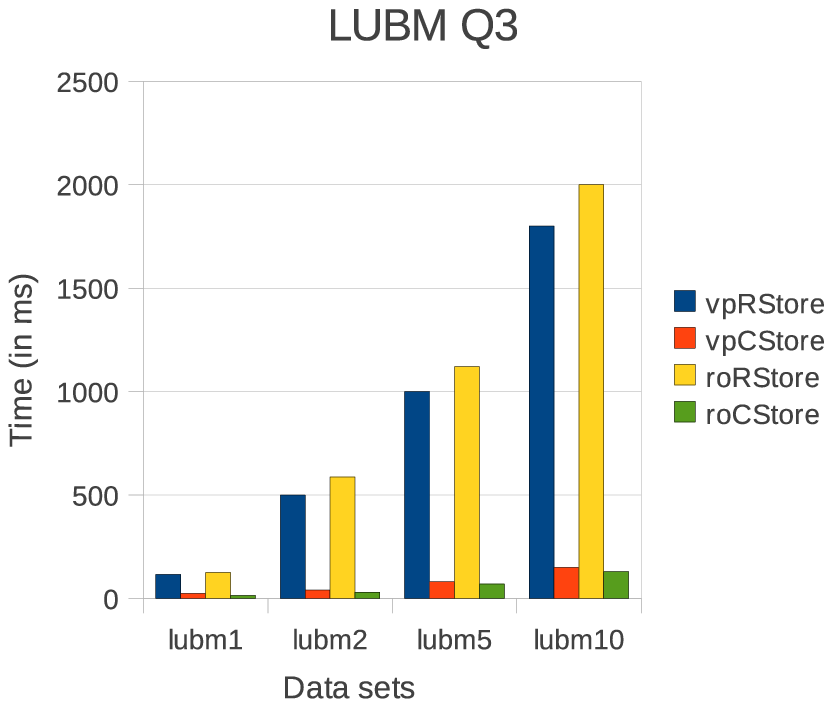}
\caption{Performance results for Q3}
\label{q3}
\end{minipage}
\hspace{0.1cm}
\begin{minipage}[b]{0.5\linewidth}
\centering
\includegraphics[scale=.6]{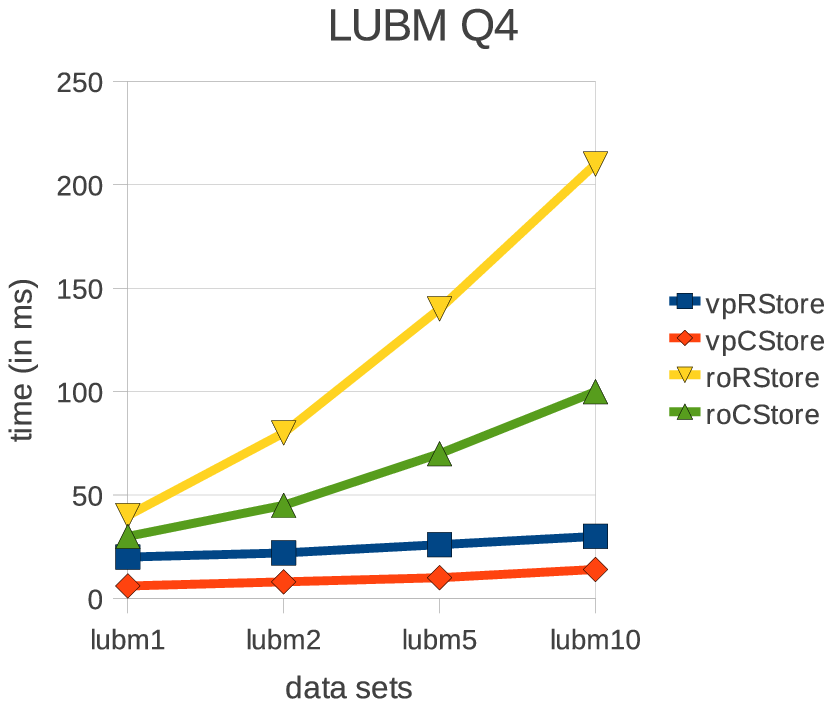}
\caption{Performance results for Q4}
\label{q4}
\end{minipage}
\end{figure}

\noindent\textbf{Analysis of Q4.} Once again, in this query, we do not use inference on the \texttt{worksFor} \textit{predicate} to include results on sub-properties of this last. 
This is motivated by the will to emphasize on the weaknesses of the \texttt{roStore} approach. As expected, \texttt{vpStore} is, in this context, outperforming \texttt{roStore}. 
In fact, even \texttt{vpRStore} is outperforming \texttt{roCStore}; which can be induced by the high selectivity nature of the query (four attributes in result set).

\noindent\textbf{Analysis of Q5.} Due to the exploitation of the sub-properties of the \textit{predicate} \texttt{memberOf} in this query, it is not surprising that
\texttt{roStore} outperforms \texttt{vpStore}. Indeed in \texttt{vpStore}, the results of the query comes from the union of three distinct queries (one for each \textit{predicates} involved) 
while \texttt{roStore} only requires a single query.
 
\noindent\textbf{Analysis of Q6.} 
This query retrieves the subjects from a two columns table (i.e. type). Because the column stores primarily order these relations on the subject, 
they are more efficient than their row store counterparts. This is due to better I/O efficiency.
 Similarly to Q3, the \texttt{roStore} approach outperforms \texttt{vpStore}.

\noindent\textbf{Analysis of Q7.} Again query processes in \texttt{roStore} (resp. column store) is more efficient than \texttt{vpStore} (resp. row store). 
The reasons are similar to the ones for Q3.

\noindent\textbf{Analysis of Q8.} The analysis of the results for this query confirm the ones of Q5.%: \texttt{roStore} outperforms \texttt{vpStore} and column stores are faster than row stores.

\begin{figure}[ht]
\begin{minipage}[b]{0.5\linewidth}
\centering
\includegraphics[scale=.6]{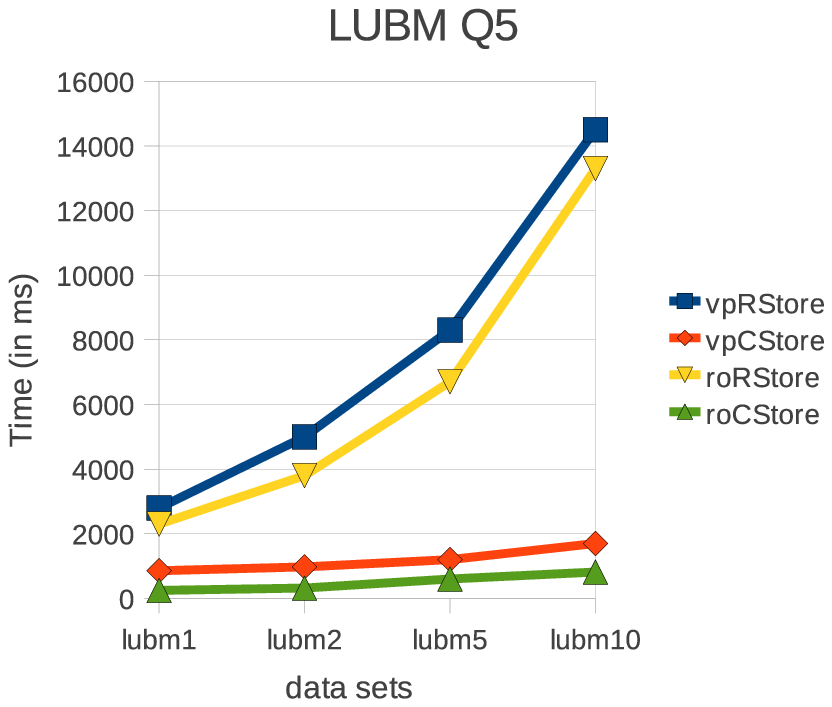}
\caption{Performance results for Q5}
\label{q5}
\end{minipage}
\hspace{0.1cm}
\begin{minipage}[b]{0.5\linewidth}
\centering
\includegraphics[scale=.6]{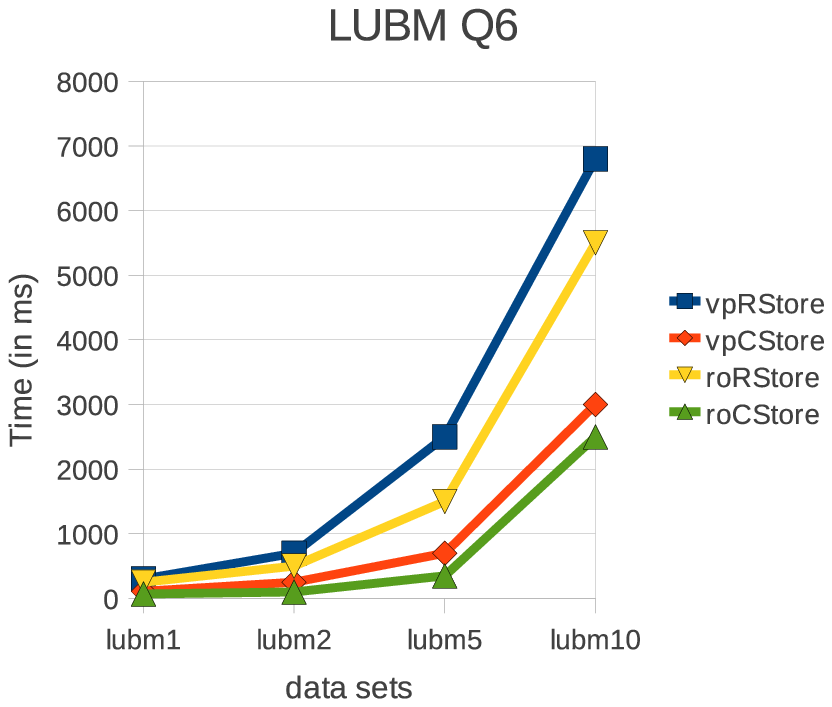}
\caption{Performance results for Q6}
\label{q6}
\end{minipage}
\end{figure}

\begin{figure}[ht]
\begin{minipage}[b]{0.5\linewidth}
\centering
\includegraphics[scale=.6]{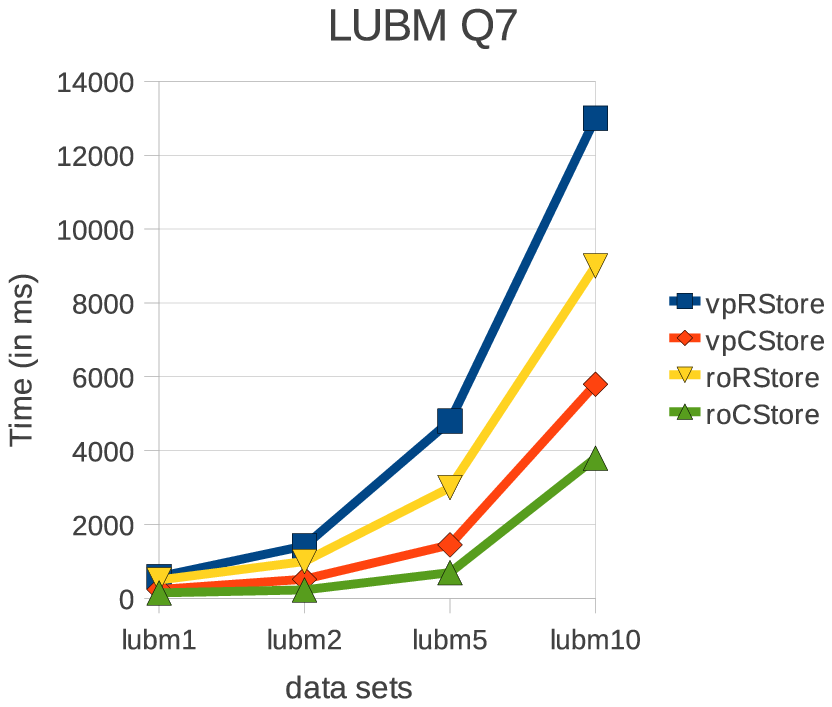}
\caption{Performance results for Q7}
\label{q7}
\end{minipage}
\hspace{0.1cm}
\begin{minipage}[b]{0.5\linewidth}
\centering
\includegraphics[scale=.6]{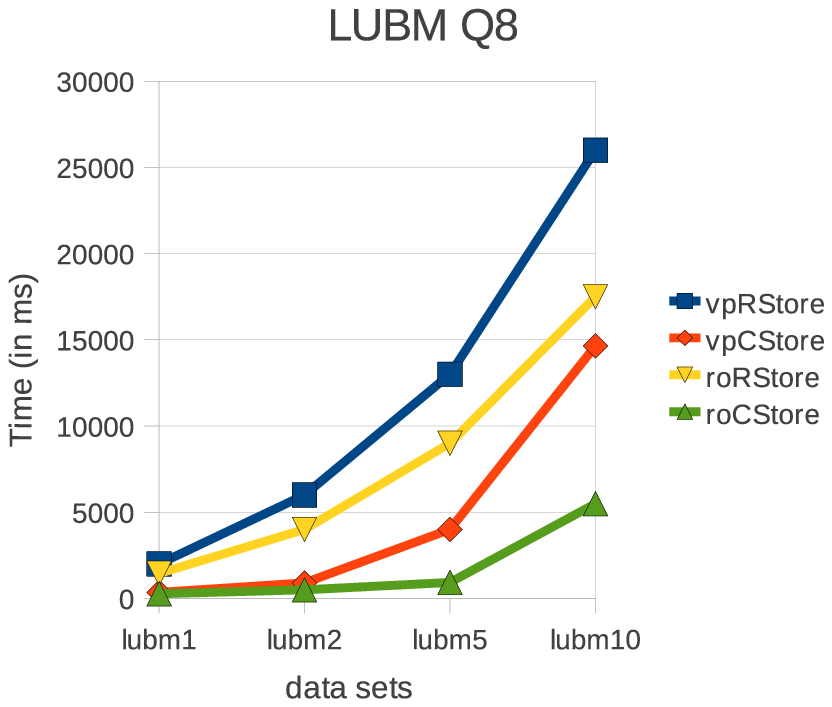}
\caption{Performance results of Q8}
\label{q8}
\end{minipage}
\end{figure}

\noindent\textbf{Analysis of Q9.} This query does not require inferences on property hierarchies but some on several concept ones. 
As seen previously, in this situation column stores is more efficient than row stores.
 On column stores, \texttt{vpStore} and \texttt{roStore} have close performance results, with \texttt{roStore} slighty better than \texttt{vpStore}.

\noindent\textbf{Analysis of Q10.} The results are similar to Q9.

\begin{figure}[ht]
\begin{minipage}[b]{0.5\linewidth}
\centering
\includegraphics[scale=.56]{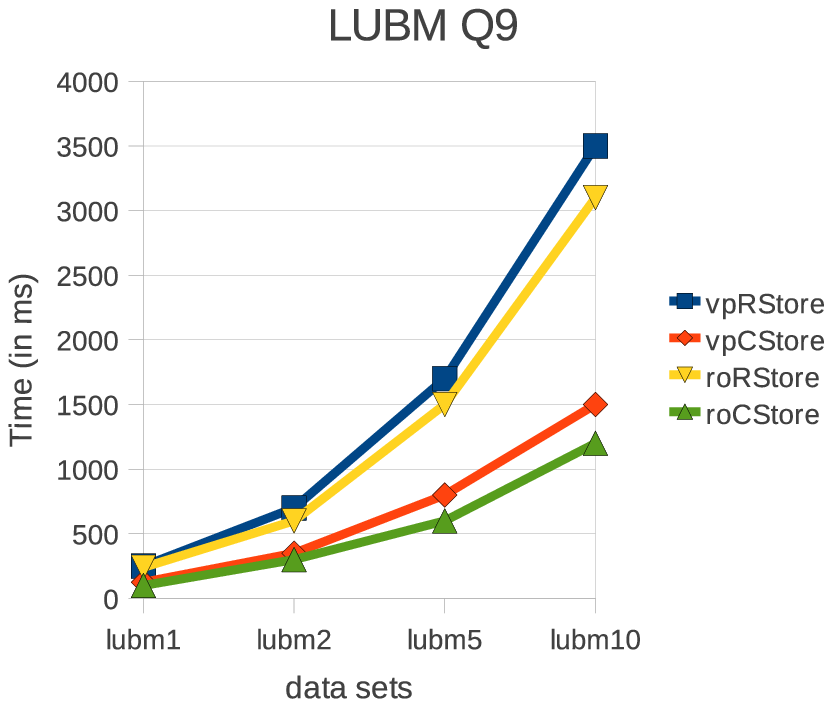}
\caption{Performance results for Q9}
\label{q9}
\end{minipage}
\hspace{0.1cm}
\begin{minipage}[b]{0.5\linewidth}
\centering
\includegraphics[scale=.56]{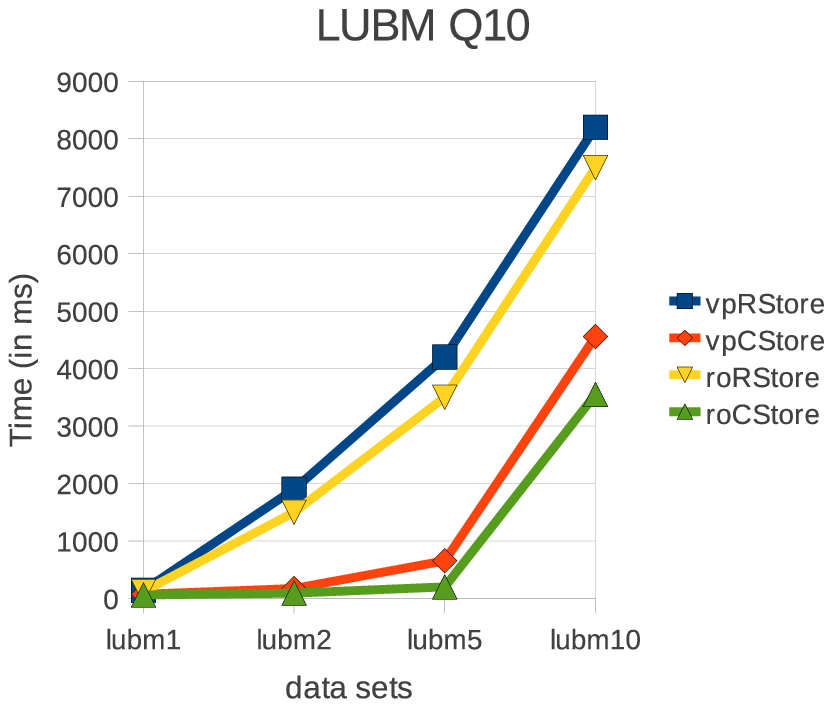}
\caption{Performance results of Q10}
\label{q10}
\end{minipage}
\end{figure}

\noindent\textbf{Analysis of Q14.} This query has large input and low selectivity with no inferences. As expected, \texttt{roCStore} is faster than \texttt{vpCstore} which is more efficient than
\texttt{roRStore}; the less effective being \texttt{vpRStore}. Note that this is due to distinguished variable being placed at the \textit{subject} position of the only triple of the
\texttt{WHERE} clause. A similar query pattern with the distinguished variable mapped to the \textit{object} position of a triple would emphasize the superiority of the \texttt{vpStore} approach.

\noindent\textbf{Analysis of Q15.} This query clearly demonstrates the efficiency of \texttt{roStore} over \texttt{vpStore}. 
Even the row oriented \texttt{roStore} outperforms the column oriented 
\texttt{vpStore}. This is due to the presence of \texttt{UNION SQL} operators in the queries executed on the \texttt{vpStore} while \texttt{roStore} only requires a complete scan of the tuples
of one table.

\begin{figure}[ht]
\begin{minipage}[b]{0.5\linewidth}
\centering
\includegraphics[scale=.56]{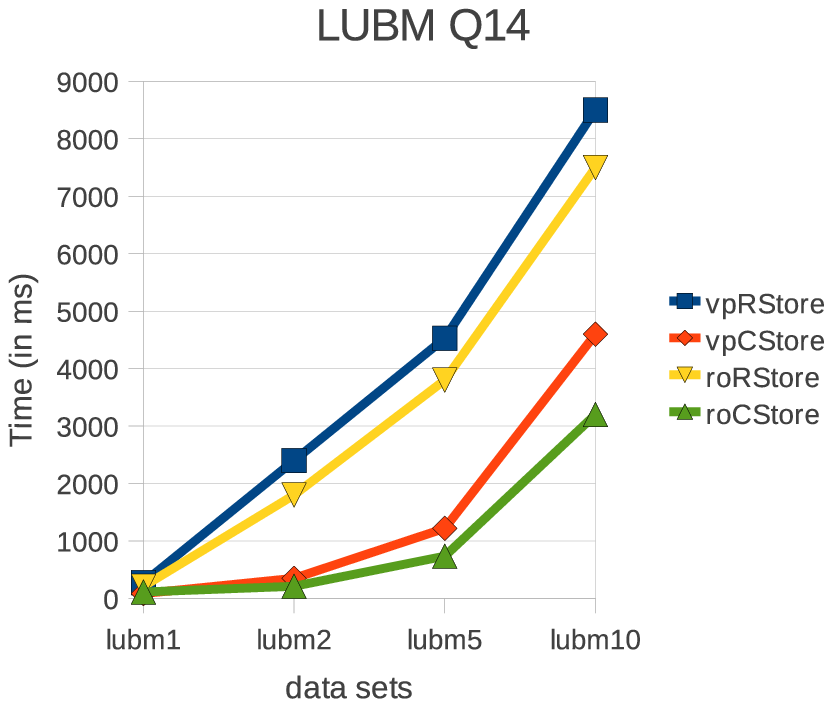}
\caption{Performance results for Q14}
\label{q14}
\end{minipage}
\hspace{0.1cm}
\begin{minipage}[b]{0.5\linewidth}
\centering
\includegraphics[scale=.56]{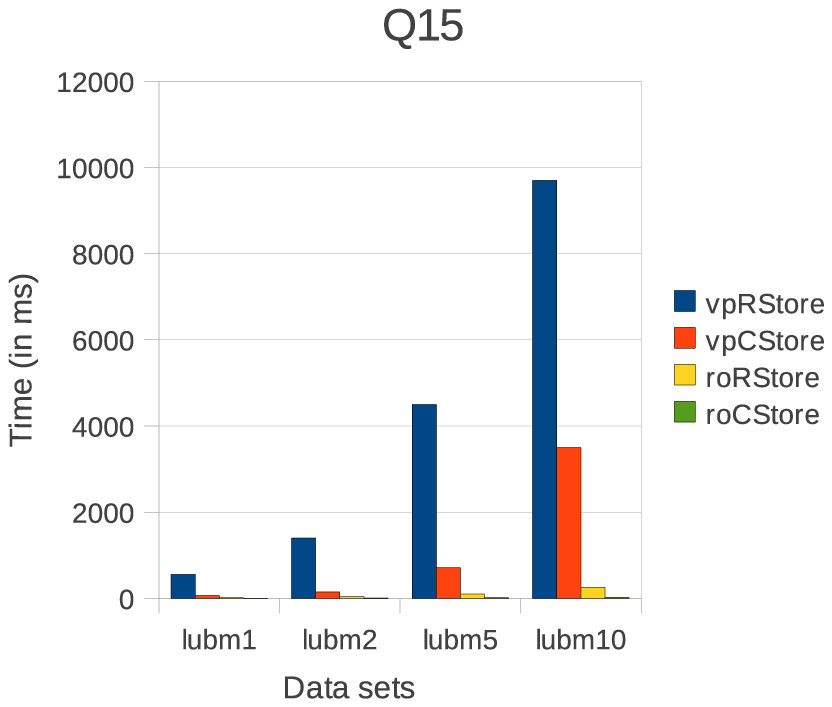}
\caption{Performance results of Q15}
\label{q15}
\end{minipage}
\end{figure}

\begin{figure}[ht]
\begin{minipage}[b]{0.5\linewidth}
\centering
\includegraphics[scale=.6]{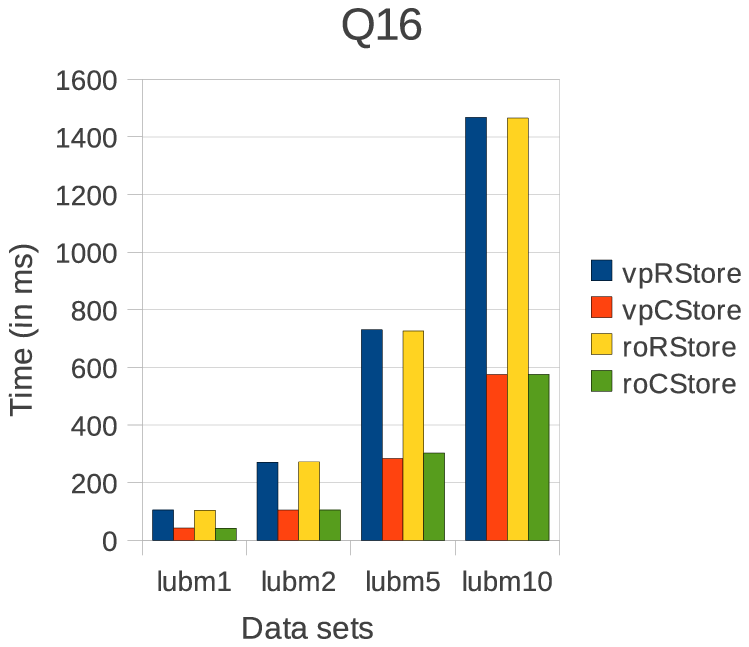}
\caption{Performance results for Q16}
\label{q16}
\end{minipage}
\hspace{0.1cm}
\begin{minipage}[b]{0.5\linewidth}
\centering
\includegraphics[scale=.6]{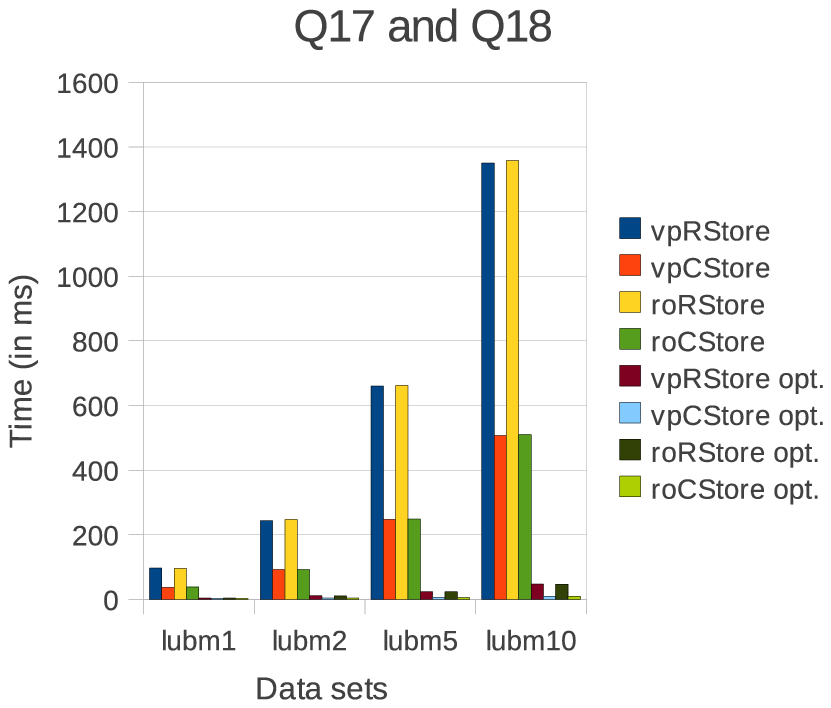}
\caption{Q17 and Q18 performance results}
\label{q17q18}
\end{minipage}
\end{figure}

\noindent\textbf{Analysis of Q16, Q17 and Q18.} Finally, queries Q16, Q17 and Q18 emphasize the importance of reasoning over the ontology before executing queries over any of the store solutions.
Figure \ref{q16} displays the duration times for all databases, ranging from approximately 42ms (column store with 1 university) to 1450ms (row store with 10 universities). This can be considered 
rather long to propose an empty answer set since, according to the ontology, the query is incoherent. Comparatively, the \texttt{propertyCheck} method we have implemented needs an average time of 
1ms to reply that the query is coherent or not. Hence, a system implemented on top of an \texttt{OWL} compliant reasoner is able to determine almost instantly
 if the answer set is empty. 

Moreover, it could also provide some explanations concerning the inconsistency of the query. We believe that such optimization are quite useful especially when end-users are not confident with all 
the details of a given ontology. The performance results of Q17 and Q18 are provided together in Figure \ref{q17q18} in order to highlight their comparisons. The purpose of Q17 and Q18 is to 
emphasize the importance of analyzing \textit{predicate} \texttt{rdfs:domain} and \texttt{rdfs:range} in a property table approach. The execution of Q17 does not perform any optimization while Q18 
checks that the concept \texttt{Faculty} is the \texttt{rdfs:domain} of the \texttt{teacherOf} \textit{predicate} and hence a join to the \texttt{rdf:type} relation is not necessary.

\textbf{Summary:}
Several conclusions can be drawn from our evaluation.
Considering the adoption of a database solution, we confirm the evaluations of \cite{Abadi} and \cite{Sidirourgos08} stating that 
column stores outperform row stores for \texttt{RDF} triple storage. The only exception in our experiments consists 
in Q2 which is rather due to the partitioning approach.

Concerning the partitioning approach, all our intuitions were confirmed by this evaluation. 
That is \texttt{roStore} outperforms \texttt{vpStore} whenever queries retrieve information from triples where properties belong a given property hierarchy
 (e.g. Q5 and Q15). 
On the contrary, \texttt{vpStore} is more efficient than \texttt{roStore} where only a subset of the properties of a property hierarchy are necessary to reply to a query 
(e.g. Q2). 
This result was expected since the \texttt{roStore} approach then requires to add additional conditions on the properties one wants to retrieve from a 'top property' relation.

Finally, the SQR approach seems to be quite useful since it does not slow down the processing of satisfiable queries
and enables to detect unsatisfiable queries efficiently (e.g. Q17 and Q18). 
Anyhow, we consider that more evaluations need to be conducted on larger ontologies to confirm these results.

\section{Conclusion}
The first contribution of this paper is to show that depending on the type of applications and queries asked to the RDF triple stores, 
different partitioning approaches can be considered. Between the two extremes of triple and vertical partitioning, we introduced the
\texttt{roStore} approach which is particularly advantageous for a certain class of queries, \textit{i.e.} those relying on deep property 
hierarchies (e.g. the OpenGalen ontology contains a property hierarchy of depth 6).
Moreover, this novel approach is a simple extension to the existing RDF column store work and can thus be easily adopted by other RDF stores.
A second contribution of this work is to propose a semantic query rewritting solution that can be adopted by most of the RDF triples we have 
presented in this paper (triples tables, vertical partitioning, \texttt{roStore}, property-class tables). This approach seems promising since 
it can be quite useful to detect unsatisfiable queries and optimizing other queries by analyzing property domains and ranges.

Our list of future works is large since we consider that several investigations need to be performed to complete the road map on efficient 
and persistent RDF triple storage. The first directions we would like to follow are ontology schema evolution in \texttt{roStore} (e.g. a 
new property hierarchy emerges or is removed from the ontology) and the consideration of concept hierarchies at the storage and querying levels.

\bibliographystyle{abbrv}
%\bibliography{roStore} 

%\bibliographystyle{abbrv}
%\bibliography{er10}

\begin{thebibliography}{10}

\bibitem{Abadi}
D.~J. Abadi, A.~M. 0002, S.~Madden, and K.~Hollenbach.
\newblock Sw-store: a vertically partitioned dbms for semantic web data
  management.
\newblock {\em VLDB J.}, 18(2):385--406, 2009.

\bibitem{Broekstra:2002:SGA:646996.711426}
J.~Broekstra, A.~Kampman, and F.~v. Harmelen.
\newblock Sesame: A generic architecture for storing and querying rdf and rdf
  schema.
\newblock In {\em Proceedings of the First International Semantic Web
  Conference on The Semantic Web}, ISWC '02, pages 54--68, London, UK, UK,
  2002. Springer-Verlag.

\bibitem{Chong:2005:ESR:1083592.1083734}
E.~I. Chong, S.~Das, G.~Eadon, and J.~Srinivasan.
\newblock An efficient sql-based rdf querying scheme.
\newblock In {\em Proceedings of the 31st international conference on Very
  large data bases}, VLDB '05, pages 1216--1227. VLDB Endowment, 2005.

\bibitem{Cure:2005:SDM:2101422.2101472}
O.~Cur{\'e}.
\newblock Semi-automatic data migration in a self-medication knowledge-based
  system.
\newblock In {\em Proceedings of the Third Biennial conference on Professional
  Knowledge Management}, WM'05, pages 373--383, Berlin, Heidelberg, 2005.
  Springer-Verlag.

\bibitem{DBLP:journals/ws/GuoPH05}
Y.~Guo, Z.~Pan, and J.~Heflin.
\newblock Lubm: A benchmark for owl knowledge base systems.
\newblock {\em J. Web Sem.}, 3(2-3):158--182, 2005.

\bibitem{3store2003}
S.~Harris and N.~Gibbins.
\newblock 3store efficient bulk rdf storage.
\newblock In {\em Proc. 1st Int. Workshop on Practical and Scalable Semantic
  Systems (PSSS'03)}, pages 1--20, Sanibel Island, FL, USA, 2003.

\bibitem{Kolas_efficientlinked-list}
D.~Kolas, I.~Emmons, and M.~Dean.
\newblock Efficient linked-list rdf indexing in parliament, 2009.

\bibitem{Sidirourgos08}
L.~Sidirourgos, R.~Goncalves, M.~L. Kersten, N.~Nes, and S.~Manegold.
\newblock Column-store support for rdf data management: not all swans are
  white.
\newblock {\em PVLDB}, 1(2):1553--1563, 2008.

\bibitem{Weiss:2008:HSI:1453856.1453965}
C.~Weiss, P.~Karras, and A.~Bernstein.
\newblock Hexastore: sextuple indexing for semantic web data management.
\newblock {\em Proc. VLDB Endow.}, 1(1):1008--1019, Aug. 2008.

\bibitem{Wilkinson03efficientrdf}
K.~Wilkinson, C.~Sayers, H.~Kuno, D.~Reynolds, and J.~Database.
\newblock Efficient rdf storage and retrieval in jena2.
\newblock In {\em EXPLOITING HYPERLINKS 349}, pages 35--43, 2003.

\end{thebibliography}
\end{document}